\documentclass[%
 reprint,
showpacs,preprintnumbers,
showkeys,
 amsmath,amssymb,
 aps,
 pre,
floatfix,
]{revtex4-2}

\usepackage{mathtools}
\usepackage{graphicx}
\usepackage{dcolumn}
\usepackage{bm}
\usepackage{hhline}
\usepackage{color,soul}
\usepackage{float}
\usepackage[normalem]{ulem}
\usepackage[dvipsnames]{xcolor}
\usepackage[sort&compress]{natbib}
\usepackage{xr-hyper}
\usepackage{hyperref}
\hypersetup{
    colorlinks,
    linkcolor={red!70!black},
    citecolor={blue!70!black},
    urlcolor={blue!70!black}
}
\usepackage{pgffor}
\usepackage{pdfpages}
\makeatletter
\AtBeginDocument{\let\LS@rot\@undefined}
\makeatother

\begin{document}


\title{Pattern stabilisation in swarms of programmable active matter: a probe for turbulence at large length scales}

\author{Pankaj Popli}
\affiliation{
Tata Institute for Fundamental Research, Centre for Interdisciplinary Sciences, 36/P Gopanapally, Hyderabad 500046, India\\
}
\author{Prasad Perlekar}
\affiliation{
Tata Institute for Fundamental Research, Centre for Interdisciplinary Sciences, 36/P Gopanapally, Hyderabad 500046, India\\
}
\author{Surajit Sengupta}\thanks{deceased}
\affiliation{
Tata Institute for Fundamental Research, Centre for Interdisciplinary Sciences, 36/P Gopanapally, Hyderabad 500046, India\\
}

\date{\today}
             
\begin{abstract}
We propose an algorithm for creating stable, ordered, swarms of active robotic agents arranged in any given pattern. The strategy involves suppressing a class of fluctuations known as ``non-affine" displacements, viz. those involving non-linear deformations of a reference pattern, while all (or most) affine deformations are allowed. We show that this can be achieved using precisely calculated, fluctuating, thrust forces associated with a vanishing average power input. A surprising outcome of our study is that once the structure of the swarm is maintained at steady state, the statistics of the underlying flow field is determined solely from the statistics of the forces needed to stabilize the swarm.
\end{abstract}

\pacs{Valid PACS appear here\\}

\maketitle
Bird flock or fish school often shows large-scale collective and coordinated motion~\cite{bird-flocks,bird-flocks2, micro-flock-1, micro-flock-2, micro-flock-3,flocks-1,flocks-2,flocks-3}. Such behavior not only protects individuals in the group from predators but also reduces hydrodynamic effects~\cite{flocks-3,advntg-schooling-2,advntg-schooling-3,sriram-1,sriram-3}. Pattern formation while flocking, of course, is an interplay between hydrodynamic interactions mediated by embedded fluid medium and how well swimmers respond to it~\cite{learning-schooling,samridhi-3,sriram-1, sriram-2,sriram-3,Tonner-2}. Studies have shown that self-propelled agents, equipped with reinforcement learning algorithms, can adapt to minimize collective flying efforts and produce stable geometrical patterns~\cite{learning-schooling}. Experiments have also used inanimate, autonomous agents such as active colloidal particles or robotic agents to mimic collective motion observed in nature~\cite{self-flock-1,self-flock-2,micro-flock-4,field-driven-bots,vicsek-1,vicsek-2}.
\par
The use of a patterned swarm of drones has shown immense potential in surveying, disaster management, and setting up a communication network in inaccessible locations~\cite{drones-1,target-localization-drone,gas-sensing-drone,patent-1,uav-wireless-network,flying-cellu-comm,project-loon,starlink,project-owl}. Typical strategies for maintaining a pattern involves accurate measurement of the velocity of the fluid medium and actively compensating for the disruptive forces at the level of individual agents using computations performed at a central command and control station~\cite{ram-ramaswamy,sudeshna-adaptive-control,control-theory}. Although a fixed patterned arrangement of drones is obtained, the resulting swarm requires calm weather to operate and uses various collision avoidance and machine-learning algorithms, which increase operational and computational complexity~\cite{learning-drones,intel-drone-show,Khurana2013,vicsek-drones,samridhi-1}.

\begin{figure}[!h]
\begin{center}
\includegraphics[scale=0.55]{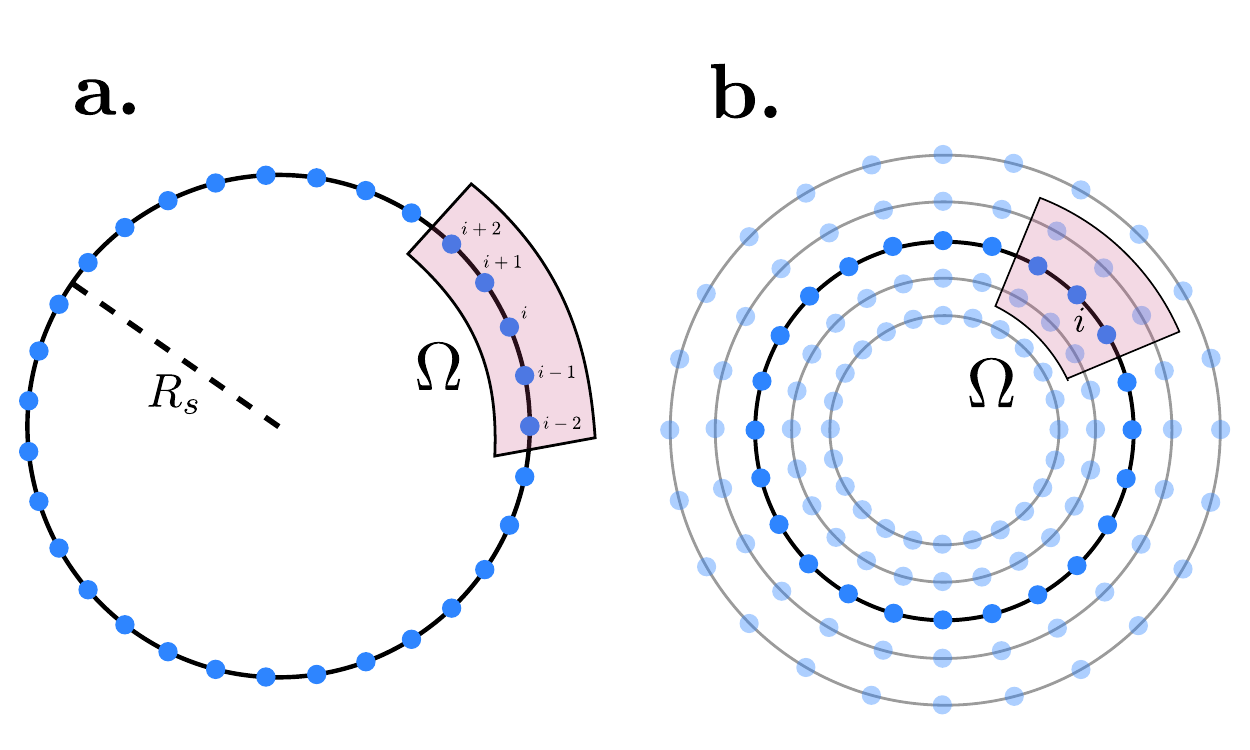}
\end{center}
\caption{\label{fig:model-a-b-omega}Reference structure and coarse-graining volume $\Omega$. {\bf a.} Model A : Floppy Swarm, Active particle spaced equally on a ring. Around particle $i$ coarse-graining volume (pink shaded) is defined and consists two left and right neighbours.{\bf b.} Model B : Rigid Swarm, Active particles equally spaced on a ring (dark blue) surrounded by layer of concentric ghost particles (light blue). Coarse-graining volume (pink shaded) around particle $i$ consists of total $8$ particles.}
\end{figure}
\par
In this letter, we propose a strategy to produce positionally ordered patterns of drones or robotic agents or ``bots''~\cite{uav-book,drones-1,intel-drone-show} that are robust, do not require velocity sensors or any a priori knowledge of the underlying flow field. Another significant finding of our work is that in certain conditions, statistical properties of the underlying flow field can be obtained, as a co-product, without using any invasive velocity probes. A common approach for studying flow statistics involves point particles that preferentially sample the flow structures or using an extended object such as a polymer if two-point correlations are needed~\cite{tracer-trblnt-1,tracer-trblnt-2,wilczek-tracer,samridhi-2,samridhi-4,samridhi-5,probe-trblnt-3,probe-trblnt-4,probe-polymer}. In atmospheric turbulence, these approaches are inefficient to obtain Eulerian statistics since particles separate quickly from one another due to turbulent diffusion~\cite{turbulence-book}. This knowledge, on the other hand, would be of great help in interpreting wind patterns~\cite{wind-pattern} and weather prediction~\cite{richardson-book}, wind energy generation~\cite{bandi-prl}, understanding storms and hurricanes etc.
\par
Consider a system of bots placed in some pattern and embedded in a flowing medium. For specificity and computational simplicity, we consider here point particles placed at equal intervals on a circular ring of radius $R_s$ in two dimensions (see Fig.~\ref{fig:model-a-b-omega}). Our analysis is general and unchanged if any other pattern or dimension is chosen. The particles have mass $m$ and are allowed to move within a square box of length $L>> R_s$. At any time  $t$, the position  $\mathbf{r}_i$ and velocity $\mathbf{v}_i$ of $i^{th}$ particle is determined by the set of Stokes-drag equations~\cite{batchelor} with a constant drag coefficient $\gamma$ and in the presence of a background flow field $\mathbf{U}(\mathbf{r}_i,t)$, viz.
 \begin{eqnarray}
\cr\frac{d\mathbf{r}_i}{dt}&=&\mathbf{v}_i\hspace{2pt},\\
\cr m\frac{d\mathbf{v}_i}{dt}&=&-\gamma\left(\mathbf{v}_i-\mathbf{U}(\mathbf{r}_i,t)\right)+\vec{\mathcal{F}}_i\hspace{2pt}, \label{eq:eqofmotion}
\end{eqnarray}
Here, $\vec{\mathcal{F}}_i$ represents active forces produced by the bots' own propulsion mechanism. These forces may, in principle be programmed to mimic intra-bot interactions arising from a virtual Hamiltonian. We put $m=1$ for our subsequent analysis. Note that, we assume that the bots are small in comparison to the typical flow structure so that their presence does not significantly alter the flow field $\mathbf{U}(\mathbf{r}_i,t)$. 
\par
Here $\mathbf{U}({\bf r}, t)$ is the solution of the Navier-Stokes equations under appropriate boundary conditions and parameters representing fully developed turbulence. To reduce computational cost, however, we use a synthetic, multi-scale, spatio-temporally correlated turbulent-like flow field as described in~\cite{Fung-1,fung-2,fung-3} for most of our results. We also show later that our main conclusions are unchanged if a realistic $\mathbf{U}({\bf r}, t)$ obtained using Direct Numerical Simulation (DNS)~\cite{perlekar2017,xiao09} is used.
\par
The velocity of the field $\mathbf{U}(\mathbf{r},t)$ at any given position and time is obtained from the Fourier series,
$\mathbf{U}(\mathbf{r},t)=V_0\sum_{n=1}^{N_k}\big[ \mathbf{A}_{n}\cos(\mathbf{k}_n\cdot\mathbf{r}+\omega_{n}t) +\mathbf{B}_{n}\sin(\mathbf{k}_n\cdot\mathbf{r}+\omega_{n}t)\big].$
Here $N_k$ is the total number of Fourier modes included and $V_0$ is a dimensionless constant which determines the strength of the field. Also, the Fourier coefficients $\mathbf{A}_n$, $\mathbf{B}_n$ and distribution of modes $\mathbf{k}_{n}$, $\omega_{n}$ are chosen randomly with the constraint that the velocity field produced is incompressible and the energy spectrum follows Kolmogorov scaling $k^{-5/3}$; see Supplemental Materials (SM)~\cite{SM}.
\par
What is the nature of the forces $\vec{\mathcal{F}}$ needed to maintain the shape of any given pattern (e.g. a ring) in the presence of turbulent field $\mathbf{U}(\mathbf{r},t)$? It is easy to see that the simplest choice viz. nearest neighbor harmonic forces~\cite{spring-turblnt-1,spring-trblnt-2,spring-trblnt-3,spring-trblnt-4,spring-trblnt-5}, will not work. Local harmonic forces alone cannot guarantee the stability of the global pattern and under the influence of a turbulent velocity field, the polymer quickly intertwines with itself and collapses~\cite{SM}. Increasing the range of the harmonic forces is ineffective and simply increases the persistence length - with intertwining happening further and further apart unless a large fraction of particles are bonded to each other~\cite{stability}. This increases the computational and communication overheads for large swarms. An alternative may be to incorporate three-body forces that prefer particular bond angles~\cite{kkbend1,kkbend2}. Even in the presence of such interactions, the desired pattern may not be the unique ground state~\cite{kedia-bistable-spring}, and formulating such interactions may become cumbersome for more complex patterns. We describe below an algorithm which solves this problem, guaranteeing  a unique global order using forces derived {\it only} from configurations of the local particle neighborhood.
\par
We begin by observing that maintaining local particle connectivity guarantees global stability of any pattern, except for overall affine deformations i.e. all translations, rotations, dilations and shears. The pattern is represented by a set of tagged/labelled reference coordinates $\{ {\bf R}_i\}$ for particles (bots) $i = 1 \dots N$. Now, any set of particle displacements $\mathbf{u}_i = {\bf r}_i(t)-{\bf R}_i$ can be projected onto orthogonal non-affine and affine subspaces~\cite{sas1,sas2} by a linear projection operator ${\mathsf P}$ that we define shortly. The non-affine part of the displacements involves particle rearrangements and changes the local connectivity of the neighborhood. We show that it is possible to determine time dependent forces $\vec{{\mathcal F}}_i(t)$ that selectively suppress non-affine displacements and therefore maintain local connectivity~\cite{poplitrap,sas4,sas7,sas3}.  Since computation of $\vec{{\mathcal F}}_i(t)$ stems from a linear optimization problem, the desired reference structure is guaranteed to be the unique ground  state~\cite{sas1,popli2019}. While detailed discussions of the non-affine projection formalism is given in~\cite{sas1,popli2019} (see also SM~\cite{SM}), we briefly recall the main ideas relevant to the present study.
\par
Around each particle $i$, define a neighborhood $\Omega(i)$ consisting of the neighbours of the $i^{th}$ particle. In $d$ dimension for a given coarse-graining volume consisting of $N_{\Omega} \leq N$ particles  there exist $d^2$ affine and $N_{\Omega}d -d^2$ non-affine displacement modes~\cite{sas1,popli2019}. Clearly, for non-affine displacement modes to exist, $N_{\Omega}$ requires to satisfy the obvious condition $N_{\Omega}>d$.
For any generic deformation of $\Omega(i)$, we construct a block column vector $\mathbf{\Delta}$ of size $N_{\Omega}d$ whose elements are $\Delta_{j\alpha} = u_j^{\alpha}-u_{i}^{\alpha}=(r_j^{\alpha}-r_{i}^{\alpha})-(R_j^{\alpha}-R_{i}^{\alpha}), \quad\forall\quad j\in\Omega(i)$, i.e. the $\alpha$-th component of the relative displacement of particles $i$ and $j$.  Next we define the linear projection operator ${\mathsf P}$ of $\mathbf{\Delta}$ onto non-affine subspace such that non-affine component of displacements is given by $\mathsf{P}\mathbf{\Delta}$ and $\chi_i(\{\mathbf{r}\},\{\mathbf{R}\}) = \mathbf{\Delta}^{\mathrm{T}}\mathsf{P}\mathbf{\Delta}$ measures total non-affinity associated with the deformed $\Omega(i)$. The projection operator $\mathsf{P}$ is a function only of the reference structure and is given by $\mathsf{P} = \mathsf{I-M(\mathsf{M^{\mathrm{T}}M})^{-1}\mathsf{M}^{\mathrm{T}} }$ with block matrix $\mathsf{M}_{j\alpha,\mu\nu} = \delta_{\alpha\mu}(R_{j}^{\nu}-R_{i}^{\nu})$, where $R_{j}^{\nu}, R_{i}^{\nu}$ are $\nu^{th}$ components of the desired reference position of active particles for $j\in\Omega(i)$~\cite{SM}. The active forces for the swarm are then defined as
$
\vec{\mathcal{F}}_{i}=-\partial(-h_{X}NX)/\partial\mathbf{r}_i
$
where the parameter $h_{X}$  determines the strength of non-affine active forces conjugate to global non-affinity field 
$
X(\{\mathbf{r}\},\{\mathbf{R}\}) = N^{-1}\sum_i\chi_i
$. Note that although $X$ is a multiparticle potential,  it is fairly short-ranged and only the neighbouring drones within coarse-grained region of a given bot indexed $i$ contributes to the gradient of $X$ (see Fig.~\ref{fig:model-a-b-omega}, Supplemental Material~\cite{SM} and Fig.~S1).   In other words, positional information of only the neighbouring drones is required to calculate non-affine forces, resulting in reduced communication and computational cost. By construction, negative values of $h_X$ selectively suppress non-affine displacements of the swarm while positive values enhance them.
For any desired reference structure, the matrix $\mathsf{P}$ is calculated and uploaded into the memory of drones. 
At any time $t$, a drone, $i$, can determine the forces $\vec{{\mathcal F}}_i(t)$ from the known instantaneous positions of its neighbors. Once applied in the form of extra thrust, these forces tend to reduce $X$, and stabilize the pattern.
\par
We introduce two distinct models viz. ``Model A" : a {\em floppy} swarm and ``Model B" : a {\em rigid} swarm, see Fig~\ref{fig:model-a-b-omega}. In Model A we use a coarse-graining volume $\Omega$ consisting of four particles viz. two left and right neighbours of the central particle. On the other hand, in Model B, active particle are sandwiched between two concentric rings of ``ghost'' particles. The position of the particles in the ghost layers are not affected by the turbulent field or non-affine forces and remain virtual, stored only in the memory of robotic agents. These virtual coordinates of ghost particles are free to translate and rotate along with particles of the actual swarm. The ghost particles in this setting satisfy rigid body constraints and therefore their positions are updated by rigid translation and rotations, $\mathbf{r}+\mathbf{V}_{com}\delta t+\Phi\delta t\hat{z}\times(\mathbf{r}-\mathbf{R}_{com})$. Here $\mathbf{R}_{com}$, $\mathbf{V}_{com}$  and $\Phi$ are the centre of mass position, velocity and average angular velocity of the actual swarm. The ghost particle positions are included in the coarse-graining volume $\Omega$ and are taken into account while calculating $\vec{{\mathcal F}}_i$ for the actual bots. Coarse-graining volume for Model B thus consist of total $8$ particles as shown in Fig~\ref{fig:model-a-b-omega}.
\begin{figure}
\includegraphics[scale=0.29]{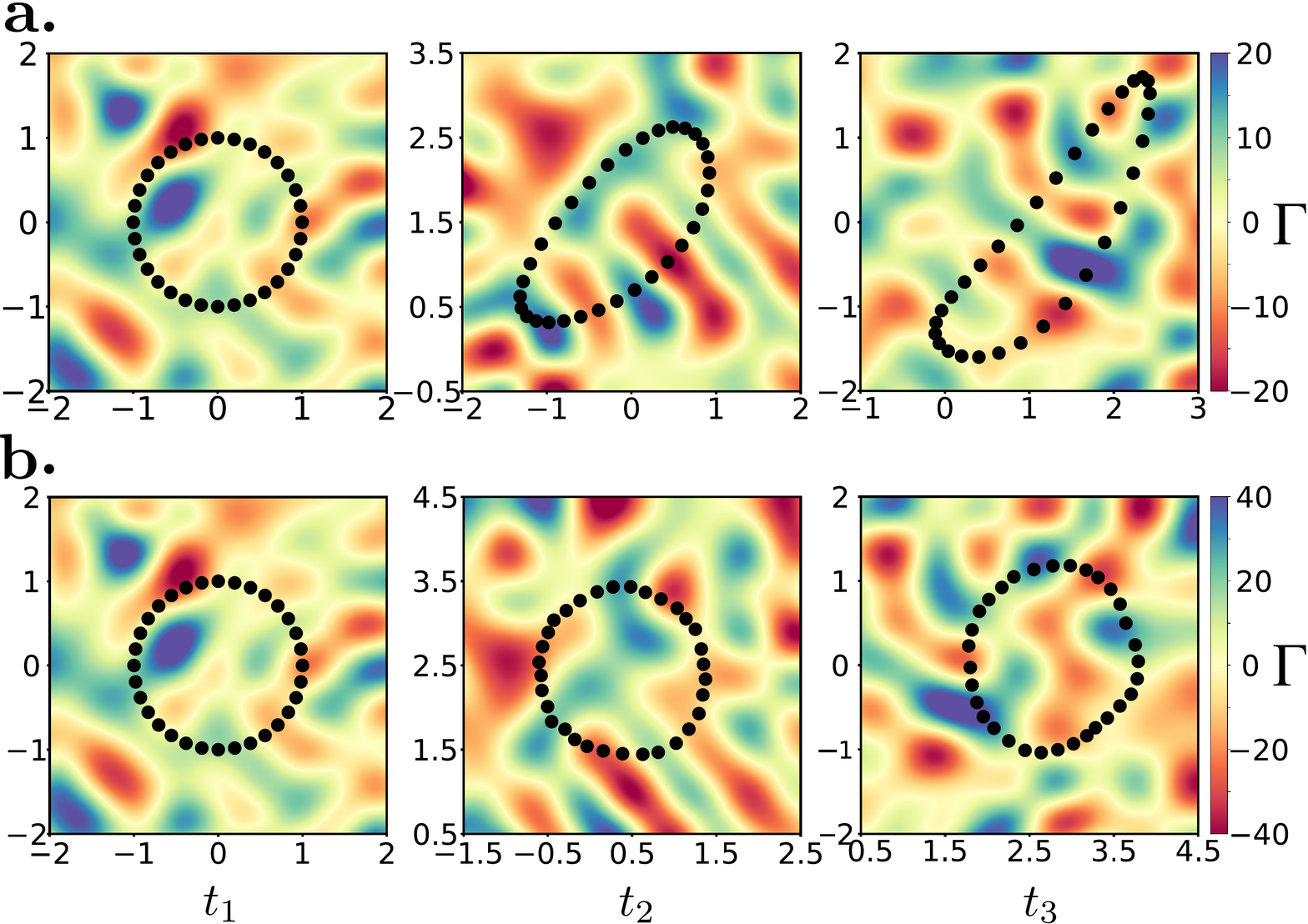}
\caption{Time evolution ($t_1<t_2<t_3$) of {\bf (a.)} Model A  ($h_{X}=-1000$,  $V_0=0.5$) and {\bf (b.)} Model B ($h_{X}=-1$,  $V_0=1$) respectively. Background color represents the vorticity values ($\Gamma=\nabla\times\mathbf{U}$) of the turbulent field. In Model A, a global transformation of the circular ring to an ellipse is allowed and cost no energy. Model B, on the other hand, is stiffer. Only a portion of the full simulation cell has been shown for clarity.}
\label{fig:movie-stills}
\end{figure}
Time dependent configurations from Model A, shown in Fig.~\ref{fig:movie-stills}{\bf a} reveals a stable elliptical pattern for sufficiently high values of $h_{X}$.  For small $h_{X}$, while the ring may occasionally intertwine with itself, it is guaranteed to disentangle back, in sharp contrast to the harmonic case. Highly elliptical configurations are, however, observed. The transformation that takes a circle to an ellipse is affine and such deformations, no matter how much large, do not produce any non-affinity in the system and are allowed as long as local neighborhood connectivity between particles is preserved. The presence of ghost particles in Model B serves as a ``stencil" for the physical swarm and all relative affine deformations of the swarm keeping ghost particles fixed are penalized. The only zero energy cost affine transformations are pure rotations and translations of the system as a whole. Fig.~\ref{fig:movie-stills}{\bf b} shows typical configurations of a Model B swarm. 
\par
\begin{figure*}
\includegraphics[scale=0.5]{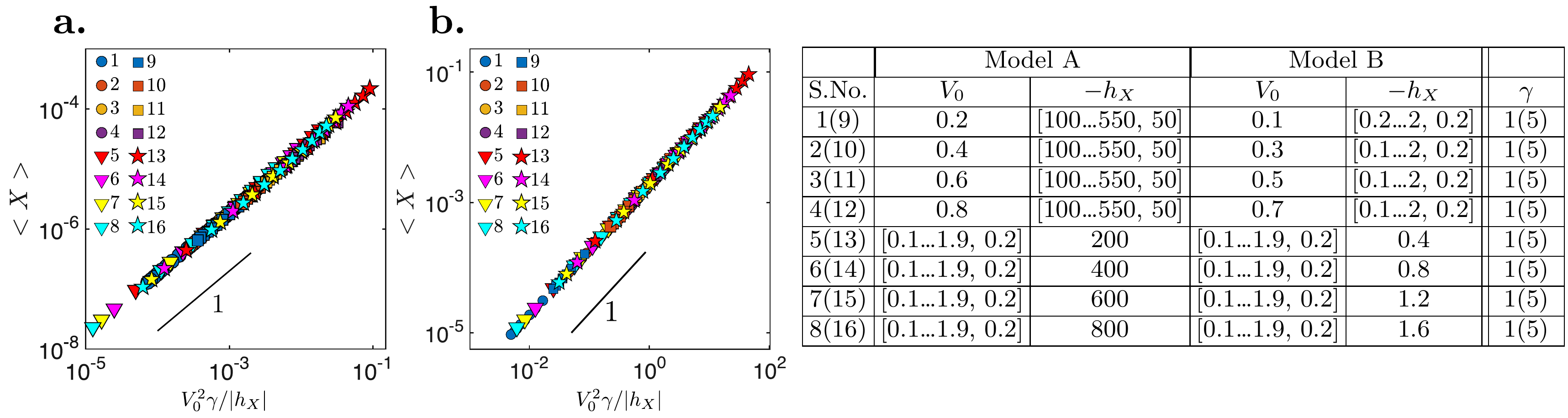}
\caption{Model A ({\bf a.}) and Model B ({\bf b.}): Global non-affinity $\langle X\rangle$ scales linearly with  $V_0^2\gamma/|h_X|$. Different symbols represents different combinations of the non-affine field $h_X$, $V_0$ and $\gamma$ (Stoke's number). Since global non-affinity measures the non-affine deformation of the swarm, an increase in the value of $X$ as a function of $V_0$ indicates the destabilizing tendency of the background flow. On the other hand, counteracting non-affine forces tend to reduce these non-affine deformations and therefore reduce $X$. Table of values of $V_0$, $h_X$ and $\gamma$ for Models A and B used for generating {\bf a}\&{\bf b} is also shown. Here the square brackets represents an interval with initial, final, and step values respectively. The parenthesis correspond to different value of $\gamma$.
}
\label{fig:x-vs-h}
\end{figure*}

It is clear that eddies with sizes much smaller than the size of the swarm cannot affect $\langle X \rangle$. On the other hand, very large eddies simply carry the swarm along introducing mainly affine deformations so that $\langle X \rangle \to 0$. Therefore $\langle X \rangle$ grows with decreasing eddy size saturating to a limit $X_c$ for eddies below a cutoff size corresponding to frequency $\omega_c$ (see SM~\cite{SM}). The only two dimensionless quantities are therefore $\langle X \rangle/X_c$ and $V_0^2 \gamma \omega_c/\vert h_X \vert$ so that the leading behavior of $\langle X \rangle \sim V_0^2 \gamma /\vert h_X \vert$ for both Models A and B as shown in Fig.~\ref{fig:x-vs-h}.
The greater flexibility of Model A implies power expended by non-affine forces is much smaller compared to B as we discuss below.
\par
Note that a constant $\langle X \rangle$ implies zero average power. Indeed, taking a dot product of both sides of the second equation in the set Eq.(\ref{eq:eqofmotion}) with the velocity $\mathbf {v}_i$ one obtains the rate of change of the kinetic energy ($=0$ in the steady state) as $\vec{{\mathcal F}}_i\cdot {\mathbf v}_i - \gamma (v_i^2 - \mathbf{U}(\mathbf{r},t)\cdot {\mathbf v}_i)$. The first term is the power expended by the non-affine forces and if ${\mathbf v}_i$ follows the local fluid velocity $ \mathbf{U}(\mathbf{r},t)$ - true for a swarm drifting along with the fluid medium, this is, on average zero. In practice, the distribution of power expended by $\vec{\mathcal F}_i$ fluctuates around zero with an error that is much smaller for Model A than for Model B as expected; see  SM~\cite{SM} for a detailed analysis for the specific case of the synthetic turbulent field used in this work.  
\par
We end this Letter by demonstrating how the Model B swarm may be used to obtain the statistics of the turbulent flow field $\mathbf{U}(\mathbf{r},t)$
\begin{figure}
\begin{center}
\includegraphics[scale=0.29,trim=0 2cm 0 4cm]{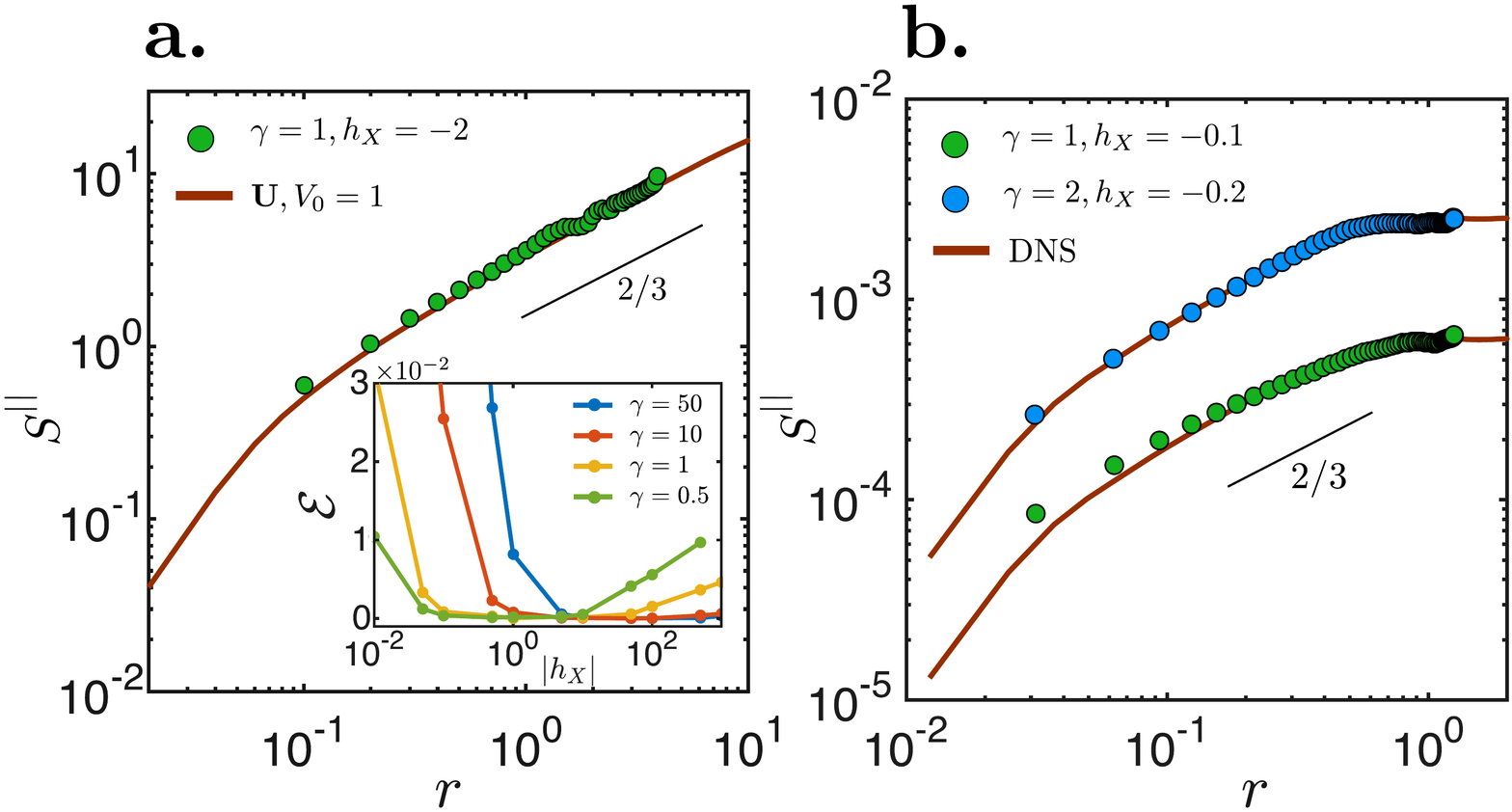}
\end{center}
\caption{Longitudinal structure function $S^{||}(r)$ as a function of distance $r=|\mathbf{r}_{ij}|$ measured at Eulerian points (solid brown line) and using non-affine forces (colored dots) in Model B. In ({\bf a.}) $\mathbf{U}({\bf r},t)$ is the same synthetic turbulent field used for earlier calculations. Simulation was done for a ring of radius $R_s=2$, box size $L=40$ and for $N=128$ particles. Inset shows linear regression error $\mathcal{E}$ of the obtained non-affine force structure function when compared to the expected $2/3$ law. Note that for a given value of $\gamma$, there exist a range of $|h_X|$ for which $\mathcal{E}$ is minimum. ({\bf b.}) Same as in ({\bf a}) but for a $\mathbf{U}({\bf r},t)$ obtained from DNS with simulation parameters as $R_s=\pi/5$, box size $L=2\pi$ and $N=128$. To highlight the identical scaling, the velocity structure function obtained using the Eulerian flow field (solid brown line) is scaled by a constant prefactor. 
}
\label{fig:stf}
\end{figure}
over a few meters to kilometers. One may be able to measure the local velocity field using appropriate sensors and compute correlation functions and structure function ``on the fly''.  
While this is of course possible, drone swarms, stabilized using forces which suppress $X$, can go a step further.  Specifically, the swarm can be used to obtain equal-time measurements over a wide range of spatial length scales.  This is useful to probe fields which involves flow structures at multiple length scales. To see this, we first point out that interestingly, no knowledge of the velocity field is necessary to derive the active forces $\vec{\mathcal F}_i$. They depend only on $\chi$ and therefore only on the knowledge of the instantaneous and reference positions of the neighbours. Obtaining information about positions is technically far easier than measuring local flow velocities.  We calculate the longitudinal structure function~\cite{turbulence-book} for non-affine forces $S_{\mathcal{F}}^{||}(r_{ij}) = \langle((\vec{\mathcal{F}}_i-\vec{\mathcal{ F}}_j)\cdot\hat{\mathbf{r}}_{ij})^2\rangle$
by binning equidistant pairs $r_{ij}$ of drones  and averaged over different realisations.
\par
Fig.~\ref{fig:stf}{\bf a} and {\bf b} shows the non-affine force structure function together with the Eulerian form computed for our synthetic turbulent field $\mathbf{U}({\bf r},t)$ and the flow field obtained from DNS(see SM~\cite{SM} for DNS details) respectively. The structure function of the non-affine forces reproduces the Eulerian curve. The presence of ghost layers, implies that all displacement of the bots except for uniform translation and rotations are considered non-affine and are counterbalanced by the restoring non-affine forces. Thus the swarm in Model B is analogous to elastic ring polymer~\cite{probe-polymer} with the rigidity parameter as $|h_X|$. However a key difference between a polymer and our swarm is that for zero or very small values of rigidity $|h_X|$, the swarm itself is not stable. The significance of $|h_X|$ on the stability of patterned swarm and the structure function is motivated by  reference~\cite{probe-polymer} and is discussed below.
\par
By looking at the momentum equation for the bots Eq.(\ref{eq:eqofmotion}), it is possible to define two characteristic time scales viz. $\tau_{\gamma}=m/\gamma$ and $\tau_{X}=\sqrt{m/|h_X|}$; associated with the damping and swarm rigidity. Depending on the ratio of these timescales, various regimes of the swarm exist. For a given background field, the value of $\tau_{\gamma}$ is set by the mass and shape of the bots. For such a system the swarm itself is not stable for too small values of $|h_X|$; in contrast to the flexible polymer described in reference~\citep{probe-polymer}. On the other hand,  a large value of rigidity ($|h_X|$) pushes normal modes of the swarm towards the higher frequencies with either little or no overlap with the frequency spectra of the turbulent field. Drawing analogy to a driven oscillator, the amplitude corresponding to driving frequencies (turbulent frequencies) becomes significantly small such that any fluctuations created by the background field dampen quickly with time scale $\tau_X$. The deformation of the swarm is therefore not subject to turbulence. This immediately suggests an optimum value of $|h_X|$ for which non-affine forces are proportional to the background velocity field and yields the desired structure function. To identify the optimum value of $|h_X|$, we plot (Fig.~\ref{fig:stf} inset) the linear regression error $\mathcal{E} = \langle(1-S_{\mathcal{F}}^{||}(r)r^{-2/3}/ \langle S_{\mathcal{F}}^{||}(r)r^{-2/3}\rangle_r)^2\rangle_r$ of the non-affine force structure function obtained compared to the predicted $2/3$ law for various values of $\gamma$ and $|h_X|$. The inset of figure~\ref{fig:stf}{\bf b} clearly shows a range of the optimum value of $|h_X|$ for which $\mathcal{E}$ is the minimum and the structure function of the non-affine forces yields the predicted $2/3$ behaviour. Therefore, an imprint of the statistics of $\mathbf{U}(\mathbf{r},t)$ is present in the statistics of the non-affine forces used to stabilize the swarm! Note that large values of $\gamma$ also requires large values of $|h_X|$ to stabilise. While of course, this is achievable in simulations, in real-world it is limited by the aerodynamics, design, and capacity of the drones to produce the required thrust.
\par
This surprising result elucidates a very important aspect of the Model B swarm. Since, non-affine forces need to be computed anyway in order to stabilize the structure, no extra measurements are required for obtaining the structure function of the background flow field. The set of robotic agents such as drones can be set up as a Model B swarm to probe the statistics of the atmospheric turbulence at large length scales. It is also easy to make a swarm switch between Model A and Model B modes because the difference is only in the reference configurations. The swarm can, therefore, fly in the floppy mode to conserve energy but switch to a stiffer configuration when a measurement of $S^{||}(r_{ij})$  is needed.
\nocite{integrator,theo-prl,pnas}.
\par
We thank S. Ganguly, P. Nath, K. Ramola, V. Pandey, and N. Rana for useful discussions.  Intra mural support from the Department of Atomic Energy, Government of India is gratefully acknowledged. 

P. Popli and P. Perlekar acknowledge the contributions of Surajit Sengupta, now sadly deceased, to the development and completion of this work.

\bibliographystyle{aipnum4-1}
%

\pagebreak
\widetext
\newpage


\section*{Supplementary material}

# Supplemental Materials
# Pattern stabilisation in swarms of programmable active matter: a probe for turbulence at large length scales

Pankaj Popli, Prasad Perlekar, and Surajit Sengupta

July 27, 2021

In these supplementary notes, we give computational details of the synthetic turbulent velocity field used, calculation of the non-affine parameter, simulation details, some additional scaling results pertaining specifically to the synthetic velocity field and details of the DNS. Results, that we expect to have general applicability have been presented in the main text.

## 1 Modelling the flow field for synthetic turbulence

We study the effect of active forces on the stabilization of a swarm of bots immersed in a turbulent flow. The concepts outlined in the paper are, however, not specific to any particular flow field or how the field is generated. For example, the turbulent field may be a solution to the Navier-Stokes equations or any model synthetic, multi-scale, spatio-temporally correlated turbulent-like flow field. For computational simplicity we choose the latter.

To model this velocity field we follow the method described in [1–3] i.e. the velocity of the field $\mathbf{U}(\mathbf{r},t)$ at any given position and time is obtained from the Fourier series,

$$\mathbf{U}(\mathbf{r},t) = V_0 \sum_{n=1}^{N_k} \left[ \mathbf{A}_n \cos(\mathbf{k}_n \cdot \mathbf{r} + \omega_n t) + \mathbf{B}_n \sin(\mathbf{k}_n \cdot \mathbf{r} + \omega_n t) \right]. \tag{S1}$$

Here $N_k$ is the total number of Fourier modes included and $V_0$ is a dimensionless constant which determines the strength of the field. Although, $\mathbf{U}(\mathbf{r},t)$ in the above equation is not a solution of Naiver-Stokes equation, the Fourier coefficients $\mathbf{A}_n$, $\mathbf{B}_n$ and distribution of modes $\mathbf{k}_n$, $\omega_n$ are chosen such that the velocity field produced is in-compressible and the energy spectrum follows Kolmogorov scaling $k^{-5/3}$. Therefore, for a $2d$ flow field, we take

$$\mathbf{A}_n = a_n(\cos(\phi_n), -\sin(\phi_n))$$

$$\mathbf{B}_n = b_n(-\cos(\phi_n), \sin(\phi_n))$$

$$\mathbf{k}_n = k_n(\sin(\phi_n), \cos(\phi_n)), \tag{S2}$$

with phases $\phi_n$ chosen randomly from a uniform distribution between $[0, 2\pi)$. The choice of phases remains constant for all time during a simulation but is different for different ensemble/realization of the field. Evidently, with the above choice of Fourier coefficients, we have $\mathbf{A_n} \cdot \mathbf{k_n} = \mathbf{B_n} \cdot \mathbf{k_n} = 0$ resulting an in-compressible field for any choice of $\phi_n$,

$$\nabla_\mathbf{r} \cdot \mathbf{U}(\mathbf{r},\mathbf{t}) = 0\,.$$

The amplitudes $a_n$ and $b_n$ in Eq.(S2) are taken,

$$|a_n|^2 = |b_n|^2 = E(k_n)\Delta k_n, \tag{S3}$$

such that $E(k)$ is the energy spectra of form $Ck^{-5/3}$ in the range $(k_1 = \frac{2\pi}{\zeta}) \leq k \leq (k_{N_k} = \frac{2\pi}{\eta})$ and zero otherwise. Here, $\zeta$ and $\eta$ corresponds to largest and smallest length scale respectively. For a discrete set of $k_n$, $\Delta k_n$ are defined as,

$$\Delta k_n = \begin{cases} (k_{n+1} - k_{n-1})/2 &, \quad 2 \leq n \leq N_k - 1 \\ (k_2 - k_1)/2 &, \quad n = 1 \\ (k_{N_k} - k_{N_k - 1})/2 &, \quad n = N_k \end{cases} \tag{S4}$$

with the wave-vector amplitudes $k_n$ obeying any one of the following distributions

$$k_n = \begin{cases} k_1 n^\alpha & \text{algebraic,} \\ k_1 \alpha^{n-1} & \text{geometric,} \\ nk_1 & \text{linear.} \end{cases} \tag{S5}$$

for $n = 1, 2, 3...N_k$. Value of $\alpha$ for the above distributions is obtained by setting $k_1$ and $k_{N_k}$ equal to $2\pi/\zeta$ and $2\pi/\eta$, thus

$$\alpha = \frac{\ln(\zeta/\eta)}{\ln(N_k)} \quad \text{for algebriac,}$$

$$\alpha = (\zeta/\eta)^{1/N_k - 1} \quad \text{for geometric.}$$

Additionally, $\omega_n$ determines the unsteadiness of the field and is proportional to the eddy turn over time, $\omega_n = \lambda\sqrt{k_n^3 E(k_n)}$, with $\lambda$ as a dimensionless constant of order 1.

In our simulations, we choose linear distribution of wave-modes with $C = 0.5$. However, any other choice of mode distribution do not affect our results qualitatively. We also set $\zeta$ equal to the length of the simulation box $L = 20R_s$ (twenty times the radius of the ring) with a total of $N_k = 500$ modes. For each realization of the flow field (choice of $\phi$), we have, in most cases, also updated the value of $\lambda$ every twentieth time step of our simulation from a Gaussian distribution of mean 1 and standard deviation 0.25 to obtain better statistics. The effect of this last step is to eliminate unnecessary oscillations arising from box commensuration effects at the cost of introducing some randomness. The frequency of updating $\lambda$ does not, in any way, affect our main results.

## 2 The non-affine projection formalism

The idea of projecting particle displacements, away from a given reference configuration, onto two orthogonal subspaces viz. affine and non-affine has been used to understand many properties of crystalline solids ranging from defects [4, 5] and pleats [6, 7] to yielding [8, 9]. In this way of classifying fluctuations, the affine part involves continuously varying displacements and can be written as a linear transformation of the reference configuration. On the other hand, the non-affine part comprises particle displacements that are not linear transformations and may not maintain topology of the local neighborhood of a particle; resulting in the nucleation of defects in crystalline solids [4, 5]. It has also been shown that a system of colloidal particles, irrespective of inter-particle interactions or noise, can be stabilized into any desired lattice symmetries if only non-affine displacements are suppressed [10]. We borrow this idea of suppressing non-affine displacements while allowing only affine transformations to stabilize patterns in

our active flocks. Below we briefly outline the projection formalism, which is utilized to stabilize ring pattern in $2d$ active robotic swarms.

Consider a system of $N$ active particles arranged in desired reference structure such as ring. We denote $\{\mathbf{u}\}$ be the set of displacement vectors away from the reference structure $\{\mathbf{R}\}$. Next, around any particular particle indexed $i$, we construct coarse-graining region $\Omega(i)$ as defined in the main text. For a pure affine deformation of $\Omega$, a deformation matrix $\mathcal{D}$ can be written such that $\mathbf{u}_j - \mathbf{u}_i = \mathcal{D}_i(\mathbf{R}_j - \mathbf{R}_i)$. On the other hand, any generic set of particle displacements contain contributions from both affine and non-affine. In such cases a local deformation tensor $\mathcal{D}$ is defined as the one which minimises

$$\chi_i = \min_{\mathcal{D}} \sum_{j \in \Omega(i)} [\mathbf{u}_j - \mathbf{u}_i - \mathcal{D}_i(\mathbf{R}_j - \mathbf{R}_i)]^2, \tag{S6}$$

the sum over index $j$ extends to all particles contained in coarse-graining volume $\Omega(i)$. In other words, $\chi$ is the measure of least square error made when particle displacements deviate substantially from those generated by a best fit affine transformation of $\Omega$. To ease the notation, we rearrange relative displacements $\mathbf{u}_j - \mathbf{u}_i$ for all $j \in \Omega(i)$ into a $2N_\Omega$ dimensional vector $\mathbf{\Delta}$ such that element $\Delta_{j\alpha}$ represents the $\alpha^{th}$ spatial component of relative displacement of particle $j$. Here $N_\Omega$ are the number of active particles in coarse-grained region $\Omega$. In similar fashion, elements of $\mathcal{D}$ are arranged into a 4 dimensional column vector $\mathbf{e} = \{\mathcal{D}_{11}, \mathcal{D}_{12}, \mathcal{D}_{21}, \mathcal{D}_{22}\}$. Thus, Eq.(S6) takes the following form,

$$\chi_i = \min_{\mathbf{e}} [\mathbf{\Delta} - \mathsf{M}\mathbf{e}]^2, \tag{S7}$$

where $\mathsf{M}$ is $2N_\Omega \times 2$ dimensional matrix with elements

$$\mathsf{M}_{j\alpha,\mu\nu} = \delta_{\alpha\mu}(R_j^\nu - R_i^\nu) \tag{S8}$$

and $R_j^\nu$ being the $\nu$th component of vector $\mathbf{R}_j$. Upon minimisation, Eq.(S7) yields

$$\chi_i = \mathbf{\Delta}^{\mathrm{T}} \mathsf{P} \mathbf{\Delta}, \tag{S9}$$

$$\mathbf{e} = \mathsf{Q} \mathbf{\Delta}.$$

Note that $\mathbf{e} = \mathsf{Q}\mathbf{\Delta}$ provides the affine deformation of $\Omega$, whereas $\mathsf{P}$ projects out non-affine part of the displacement vector $\mathbf{\Delta}$. Interestingly, the projection matrix $\mathsf{P}$ and $\mathsf{Q}$ is only a function of the desired reference structure (ring), i.e.

$$\mathsf{P} = \mathsf{I} - \mathsf{M}\mathsf{Q} \tag{S10}$$

$$\mathsf{Q} = (\mathsf{M}\mathsf{M}^{\mathrm{T}})^{-1}\mathsf{M}^{\mathrm{T}}.$$

With $\chi_i$ being the measure of non-affinity associated with particle $i$ and $\Omega(i)$, we define global non-affinity parameter for our system as $X = 1/N \sum_i \chi_i$ with $N$ being the total number of bots in the system. Higher value of $X$ corresponds to larger non-affine deformation of the coarse-graining region. In order to minimise $X$, required non-affine force of strength $h_X$ for the robotic agent $i$ is $\vec{\mathcal{F}}_i = -\partial(-h_X N X)/\partial \mathbf{r}_i$. In component form,

$$\vec{\mathcal{F}}_i = \frac{\partial}{\partial \mathbf{r}_i} \left[ h_X \sum_i^N \sum_{j,k \in \Omega(i)} (\mathbf{u}_j - \mathbf{u}_i)^{\mathrm{T}} \mathsf{P}_{j-i,k-i}(\{\mathbf{R}_i\}) (\mathbf{u}_k - \mathbf{u}_i) \right]. \tag{S11}$$

Forces represented in this form clearly show that no knowledge of the flow field or inter-particle interaction is required to stabilise the pattern. Further, desired structure of the swarm can be changed at will by

3

changing the projection matrix P. Alternatively, one can also express forces in terms of the deformation matrix $\mathcal{D}$, therefore, $\alpha^{th}$ component of the force on particle $i$ is

$$\mathcal{F}_i^\alpha = 2h_X \sum_{j \in \Omega(i)} \left[ 2u_{ij}^\alpha - \left(\mathcal{D}_i^{\alpha 1} + \mathcal{D}_j^{\alpha 1}\right) R_{ij}^1 - \left(\mathcal{D}_i^{\alpha 2} + \mathcal{D}_j^{\alpha 2}\right) R_{ij}^2 \right]. \tag{S12}$$

Here $\mathcal{D}_i$ is the best fit deformation matrix whose elements are given by,

$$\mathcal{D}_i^{\mu\nu} = \sum_{j \in \Omega(i)} u_{ij}^\mu \left(R_{ij}^1 (\mathsf{Y}_j^{-1})^{\nu 1} + R_{ij}^2 (\mathsf{Y}_j^{-1})^{\nu 2}\right)$$

$$\mathsf{Y}_j^{\mu\nu} = \sum_{k \in \Omega(j)} R_{jk}^\mu R_{jk}^\nu \tag{S13}$$

Although the force expression outlined in above is true for both Model A and B, the numerical value is different due to the sum over different coarse-grained regions and reference structures.

For completeness, we explicitly show below how to construct basic quantities such as $\boldsymbol{\Delta}$, M and projection operator P for both the models.

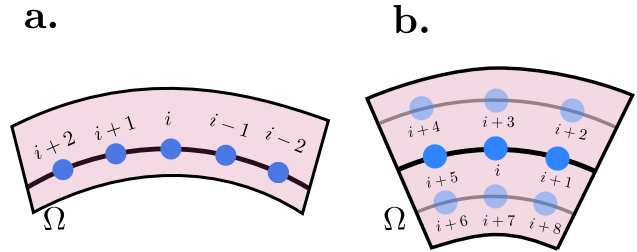

Figure S1: Coarse-graining volume $\Omega(i)$ (pink shaded) around particle $i$ for Model A (**a.**) and Model B (**b.**). Note that in Model B, although the coordinates of the ghost layer particles (transparent blue) remain virtually stored in the memory of robotic agents are considered in the reference structure and the calculation of non-affine forces.

Around any particle $i$ in Model A (see Fig. S1a), coarse-graining volume $\Omega(i)$ consists of four neighbour particles, i.e. $\Omega(i) = \{i+2, i+1, i-1, i-2\}$ (see main text). Next, for all $j \in \Omega(i)$ arrange the elements of the relative displacement $u_j^\nu - u_i^\nu$ into a $8 \times 1$ column vector to obtain

$$\boldsymbol{\Delta} = \begin{bmatrix} u_{i+2}^1 - u_i^1 \\ u_{i+2}^2 - u_i^2 \\ u_{i+1}^1 - u_i^1 \\ u_{i+1}^2 - u_i^2 \\ \vdots \\ u_{i-2}^1 - u_i^1 \\ u_{i-2}^2 - u_i^2 \end{bmatrix}.$$

The rectangular matrix M is a $4 \times 2$ block matrix with each block being a $2 \times 2$ square matrix consisting

4

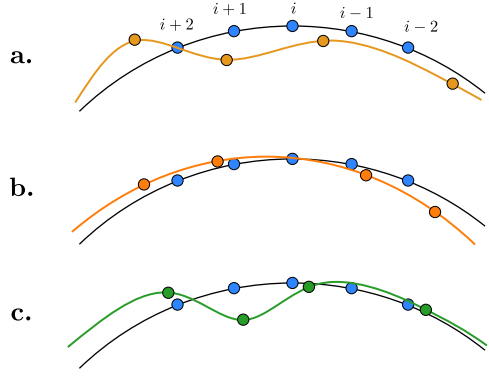

Figure S2: Model A: (**a.**) Reference (blue) and displaced (golden) positions of particles with respect to the central particle $i$. Correspond to the deformation in (**a.**), affine (orange) and non-affine (green) component of the displacements is shown in (**b.**) and (**c.**) respectively. Solid lines are for visual guidance.

only the reference position of the robotic agents. Reading from equation Eq.(S8),

$$\mathsf{M} = \begin{bmatrix} \begin{pmatrix} R^1_{i+2} - R^1_i & 0 \\ 0 & R^1_{i+2} - R^1_i \end{pmatrix} & \begin{pmatrix} R^2_{i+2} - R^2_i & 0 \\ 0 & R^2_{i+2} - R^2_i \end{pmatrix} \\ \begin{pmatrix} R^1_{i+1} - R^1_i & 0 \\ 0 & R^1_{i+1} - R^1_i \end{pmatrix} & \begin{pmatrix} R^2_{i+1} - R^2_i & 0 \\ 0 & R^2_{i+1} - R^2_i \end{pmatrix} \\ \vdots & \vdots \\ \begin{pmatrix} R^1_{i-2} - R^1_i & 0 \\ 0 & R^1_{i-2} - R^1_i \end{pmatrix} & \begin{pmatrix} R^2_{i-2} - R^2_i & 0 \\ 0 & R^2_{i-2} - R^2_i \end{pmatrix} \end{bmatrix}$$

With $\mathsf{M}$ at hand, the projection operator $\mathsf{P}$ and $\mathsf{Q}$ can be obtained easily using Eq.(S10). The projection matrix $\mathsf{P}$, once calculated, can be uploaded into the memory of robotic agents. Given any displacements of the neighbours, $\boldsymbol{\Delta}$, robotic agents then calculate non-affine forces on the fly to minimise non-affine deformations while simultaneously allowing affine displacements (see Fig. S2).

In similar fashion, for Model B, coarse-graining volume consists of total 8 particles such that $\Omega = \{i+1, i+2, i+3, i+4, i+5, i+6, i+7, i+8\}$. The displacement vector $\boldsymbol{\Delta}$ and $\mathsf{M}$, therefore, takes the following form,

$$\boldsymbol{\Delta} = \begin{bmatrix} u^1_{i+1} - u^1_i \\ u^2_{i+1} - u^2_i \\ u^1_{i+2} - u^1_i \\ u^2_{i+2} - u^2_i \\ \vdots \\ u^1_{i+8} - u^1_i \\ u^2_{i+8} - u^2_i \end{bmatrix}.$$

,

$$\mathsf{M} = \begin{bmatrix} \begin{pmatrix} R^1_{i+1} - R^1_i & 0 \\ 0 & R^1_{i+1} - R^1_i \end{pmatrix} & \begin{pmatrix} R^2_{i+1} - R^2_i & 0 \\ 0 & R^2_{i+1} - R^2_i \end{pmatrix} \\ \begin{pmatrix} R^1_{i+2} - R^1_i & 0 \\ 0 & R^1_{i+2} - R^1_i \end{pmatrix} & \begin{pmatrix} R^2_{i+2} - R^2_i & 0 \\ 0 & R^2_{i+2} - R^2_i \end{pmatrix} \\ \vdots & \vdots \\ \begin{pmatrix} R^1_{i+8} - R^1_i & 0 \\ 0 & R^1_{i+8} - R^1_i \end{pmatrix} & \begin{pmatrix} R^2_{i+8} - R^2_i & 0 \\ 0 & R^2_{i+8} - R^2_i \end{pmatrix} \end{bmatrix}.$$

Since ghost layers do not deform in Model B, any displacements of the actual swarm particles is non-affine and therefore suppressed by non-affine forces. For generalisation to more complex patterns in two or three dimensions, we refer the reader to references [5, 11].

## 3 Details of the Simulations

The equations of motion Eq.(2) in the presence of flow field and active forces are integrated numerically using a Verlet type algorithm as prescribed in [12].

$$\mathbf{r}_i(t_{n+1}) = \mathbf{r}_i(t_n) + g\delta t \mathbf{v}_i(t_n) + \frac{g\delta t^2}{2m}\mathbf{f}_i(t_n) \quad \text{(S14)}$$

$$\mathbf{v}_i(t_{n+1}) = h\mathbf{v}_i(t_n) + \frac{\delta t}{2m}\left(h\mathbf{f}_i(t_n) + \mathbf{f}_i(t_{n+1})\right),$$

where,

$$g \equiv \frac{1}{1+\gamma\delta t/2m}; \quad h \equiv \frac{1-\gamma\delta t/2m}{1+\gamma\delta t/2m}$$

Thus the particles' position $\{\mathbf{r}(t_{n+1})\}$ and velocities $\{\mathbf{v}(t_{n+1})\}$ at time $t_{n+1}$ are computed from those at the earlier time ($t_n = t_{n+1} - \delta t$). Besides that, at each time particle position and velocities are also influenced by non-affine forces (Eq.(S12)) and the background turbulent field. Therefore in Eq.(S14), $\mathbf{f}_i(t_n)$ is the total force and is given by the sum $\gamma \mathbf{U}(\mathbf{r}_i(t_n), t_n) + \vec{\mathcal{F}}_i(\{\mathbf{r}(t_n)\})$.

In our simulation, we choose $m = 1$, time step $\delta t = 0.001$ and a total of $10^5$ integration steps were performed. We check the robustness of our stabilization algorithm using the collective scalar variable $X$ averaged over different realizations of the flow field. In particular, we examine the variation of $\langle X \rangle$ with respect to various strength of the field $V_0$ and non-affine force $h_X$ and $\gamma$.

## 4 The harmonic chain in a synthetic turbulent flow

Study of a spring-mass like system immersed in a turbulent flow field has given us much insights [13–17]. Studies have shown, a spring-mass chain system can preferentially sample the vortex flow. Different motions of the spring-mass chain, such as different regimes of "flapping" motions can yield valuable information about the flow field itself such as the scaling of the velocity structure function, energy flux of the turbulent fluid, characteristic times of eddies etc [18–20]. Therefore, it is natural to ask about the stability of the patterned swarm when virtual harmonic interactions are assumed within drones. To investigate, we first consider our model swarm in the shape of a ring of particles with forces given by,

$$\mathcal{F}_i^\alpha = -K \sum_j \left(|\mathbf{r}_{ij}| - |\mathbf{R}_{ij}|\right) \frac{r_{ij}^\alpha}{|\mathbf{r}_{ij}|}, \quad \text{(S15)}$$

where sum over $j$ is extended over the neighbours of $i$ and $\alpha$ represents the spatial coordinate. In above, $\mathbf{R}_{ij}$ represents the desired relative position between particles $i$ and $j$. Although, harmonic forces used

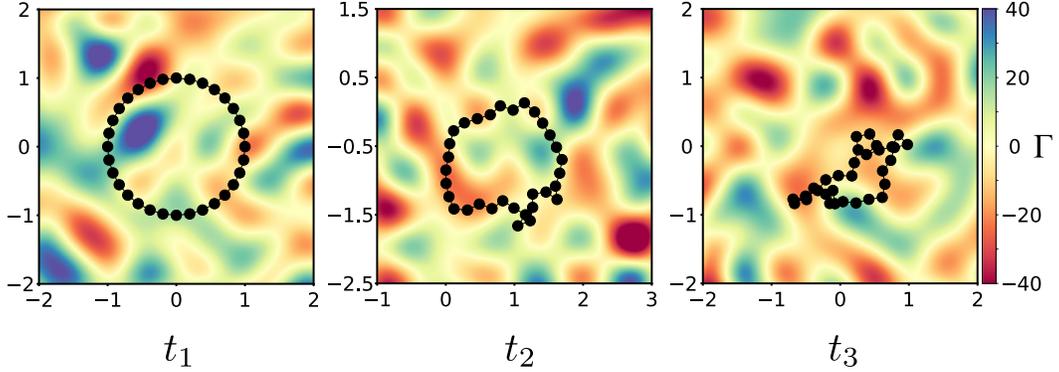

Figure S3: Typical time evolution ($t_1 < t_2 < t_3$) of the ring in the presence of turbulent field and active harmonic forces. Configurations obtained are for turbulent field strength $V_0 = 1.1$ and force stiffness $K = 500$. Background color represents the vorticity values of the turbulent field. Time $t_1$ show the initial configuration of the system. Clearly, the ring pattern of active particles is not stable if harmonic interactions are considered. Black solid lines are for visual guidance.

above are translationally invariant and depends only on the positions of neighbours, we note that such system when immersed in turbulent flow do not maintain its pattern regardless the value of stiffness $K$. Time evolution in Fig S3 shows deformation of the ring into an entangled state. Since the number of intertwined configuration far exceeds those without and the equations of motion do not explicitly prevent entanglement, large time simulation shows no recovery from this intertwined state of the ring.

## 5 Dimensional analysis and scaling.

The dimensions of each of the parameters in our system (swarm plus velocity field) are as follows,

- Turbulent field strength $[V_0] = 1$

- Energy per unit mass per unit mode $[E_n] = L^3 T^{-2}$

- Wave numbers $[k] = L^{-1}$

- Since we have $E_n = C k_n^{-5/3}$ dimensional compatibility requires $[C] = L^{4/3} T^{-2}$

- The Fourier coefficients $A_n^2 = B_n^2 = E_n \Delta k_n$ thus provides $[A_n] = [B_n] = L T^{-1}$

- Angular frequency $\omega_n = \lambda \sqrt{k_n^3 E_n} = \lambda \sqrt{k_n^3 C k_n^{-5/3}}$ then gives $[\omega] = T^{-1}$ with $[\lambda] = 1$

7

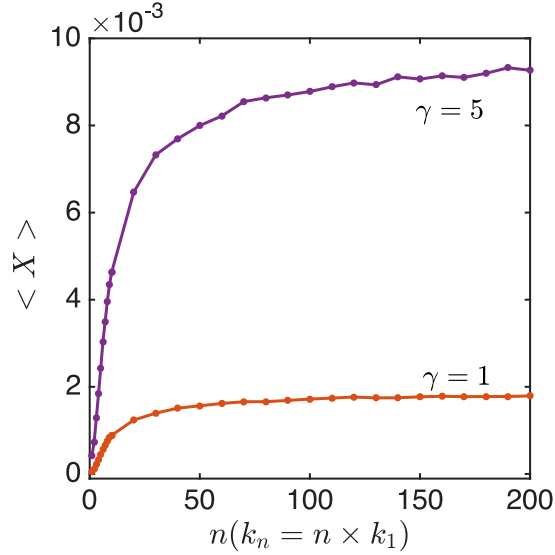

Figure S4: Variation of $\langle X \rangle$ with respect to number of Fourier modes $n$ included in the turbulent field. This plot is obtained for a Model B ring with radius $= 1$, $N = 32$ particles, $(h_X, V_0) = (-1, 1)$, and two values of $\gamma$. Total 50 ensembles are used for averaging. The plot for Model A is similar.

- non-affine field strength $[h_X] = MT^{-2}$

- the coupling constant between the field and particles $[\gamma] = MT^{-1}$.

- non-affine field strength $[h_X] = MT^{-2}$.

- Global non-affinity parameter $[X] = L^2$.

In addition, as explained in the main text, $\langle X \rangle$ approaches saturation at $X_c$ as the number of modes $n$ in the synthetic turbulent field is increased up to a cutoff $\omega_c$ (see Fig. S4) where the corresponding length scale $k_c^{-1} = L/(2\pi n)$ is much smaller than the size of the swarm. One may therefore construct only two independent dimensionless parameters $\frac{\langle X \rangle}{X_c}, \frac{V_0^2 \gamma \omega_c}{h_X}$ leading to the expectation,

$$\frac{\langle X \rangle}{X_c} = f\left(\frac{V_0^2 \gamma \omega_c}{h_X}\right),$$

which at leading order implies $\frac{\langle X \rangle}{X_c} \sim V_0^2 \gamma / h_X$. Next to leading order corrections can also be derived easily for the synthetic turbulent field as follows. Define a velocity scale such that

$$V^* = V_0 \sqrt{\frac{2}{3} \int_{k_0}^{k_c} E(k) dk} = V_0 \sqrt{\frac{2}{3} C \int_{k_0}^{k_c} k^{-5/3} dk}, \tag{S16}$$

evaluated from a lower critical value $k_0 = 2\pi n_0/L$ to $k_c = 2\pi n/L$. Evaluating the integral yields

$$(V^*)^2 = V_0^2 C \left(\frac{k_0}{n_0}\right)^{-2/3} \left(1 - \left(\frac{2\pi n}{n_0}\right)^{-2/3}\right) \tag{S17}$$

8

Finally, exchanging $V^*$ for $V_0$ in the scaling relation we get,

$$\frac{\langle X \rangle}{X_c} = \frac{V_0^2 \gamma \omega_c}{h_X}\left(1-\left(\frac{2\pi n}{n_0}\right)^{-2/3}\right) \tag{S18}$$

A direct verification of this result is shown in Fig. S5

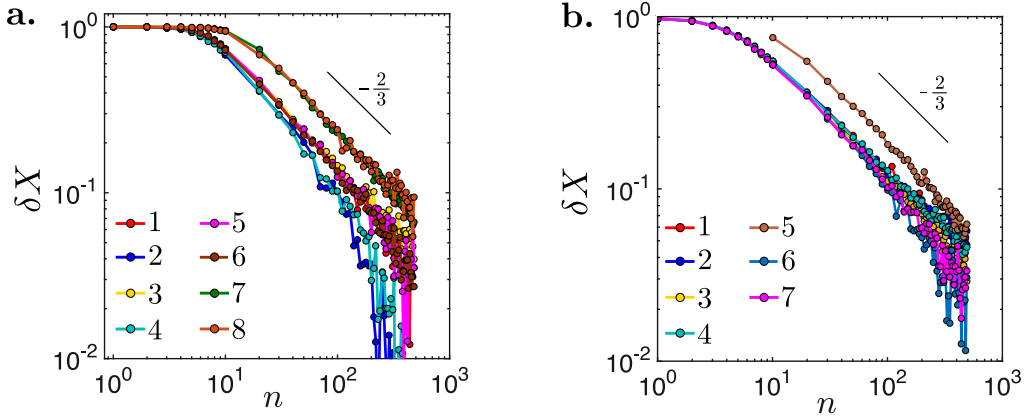

Figure S5: Variation of $\delta X \equiv 1- <X> h_X/(V_0^2 \gamma \omega_c X_c)$ with respect to number of Fourier modes $n$ included in turbulent field indicates a power-law decrease with exponent $-2/3$ for both model-A (**a.**) and model-B (**b.**). Different symbols represents different combinations of the non-affine field $h_X$, turbulent field strength $V_0$, $\gamma$ (Stoke's number), box length $L$, radius $r$, and number of bots $N$ as given in Table SI.

|  | Model A | | | | | | | Model B | | | | | | |
|---|---|---|---|---|---|---|---|---|---|---|---|---|---|---|
| S.No. | $V_0$ | $-h_X$ | $\gamma$ | C | $L$ | $r$ | $N$ | $V_0$ | $-h_X$ | $\gamma$ | C | $L$ | $r$ | $N$ |
| 1 | 0.2 | 600 | 1 | 0.5 | 20 | 1 | 32 | 1 | 1 | 1 | 1 | 20 | 1 | 32 |
| 2 | 0.6 | 600 | 1 | 0.5 | 20 | 1 | 32 | 1 | 1 | 1 | 0.5 | 20 | 1 | 32 |
| 3 | 0.2 | 800 | 1 | 0.5 | 20 | 1 | 32 | 1 | 1 | 5 | 1 | 20 | 1 | 32 |
| 4 | 0.6 | 800 | 1 | 0.5 | 20 | 1 | 32 | 1 | 3 | 2 | 0.5 | 20 | 1 | 32 |
| 5 | 0.2 | 600 | 5 | 0.5 | 20 | 1 | 32 | 1 | 1 | 1 | 1 | 40 | 1 | 32 |
| 6 | 0.2 | 600 | 1 | 1 | 20 | 1 | 32 | 1 | 1 | 1 | 1 | 40 | 2 | 32 |
| 7 | 0.2 | 800 | 1 | 1 | 40 | 1 | 32 | 1 | 1 | 1 | 1 | 40 | 2 | 128 |
| 8 | 0.2 | 800 | 1 | 1 | 40 | 2 | 128 | – | – | – | – | – | – | – |

Table SI: Table of values of $V_0$, $h_X$, $\gamma$, $L$, $r$, $C$, and $N$ for Models A and B used for generating Fig. S5

## 6 Kinetic energy and power dissipation in the steady state

As explained in the main text, multiplying equations of motion (Eq.(2) of main text) with the bot velocity yields a sum of individual power from non-affine field $\vec{\mathcal{F}}_i \cdot \mathbf{v}_i$, turbulent velocity field $\mathbf{U}(\mathbf{r}_i, t) \cdot \mathbf{v}_i$ and the self term $v_i^2$. In Fig S6, we plot histograms of $\vec{\mathcal{F}}_i \cdot \mathbf{v}_i$, the contribution to the total power from non-affine forces for both the models. For Model A, the power expended by non-affine forces also has a standard deviation that is much smaller than that of Model B.

9

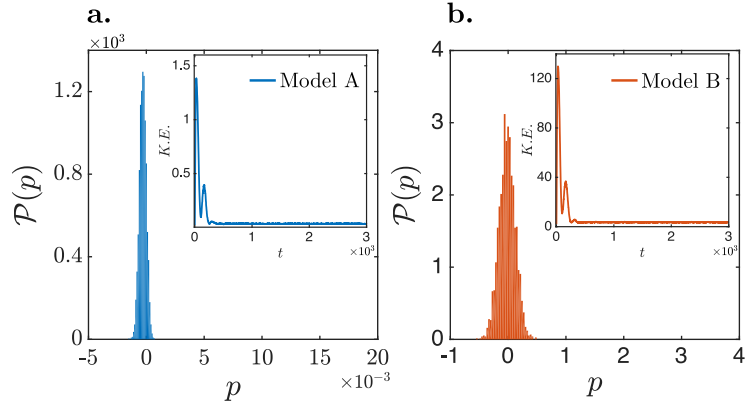

Figure S6: Normalised histograms of the contribution of non-affine forces to the power $p$ obtained for parameters $h_X = -1000, V_0 = 0.1, \gamma = 1$ for Model A (**a.**) and $h_X = -2, V_0 = 1, \gamma = 1$ B (**b.**). First each power term at any given time is averaged over 200 ensembles and then the histogram is obtained of this time series data. Note that the non-affine power fluctuates around zero. For **a** the (mean, standard deviation) of the histogram is $(-0.24 \times 10^{-3}, 0.2963 \times 10^{-3})$, for **b** the corresponding numbers are $(-0.0013, 0.14)$. Inset shows total kinetic energy of the swarm attains steady state.

# 7 Two-dimensional turbulence: Direct Numerical Simulation (DNS)

We model two-dimensional turbulent flow using incompressible, hyper-viscous Navier-Stokes equations [21, 22]:

$$\partial_t \Gamma + \boldsymbol{U} \cdot \nabla \Gamma = -\nu_i \nabla^{-4} \Gamma - \nu_u \nabla^{16} \Gamma + f_\Gamma, \tag{S19}$$

where $\Gamma = \hat{z} \cdot \nabla \times \boldsymbol{U}$ is the vorticity field, $\nu_i$ is the hypo-viscosity, $\nu_u$ is the hyper-viscosity, $f_\Gamma = f_0 \hat{\Gamma}_{\boldsymbol{k}} / \sum_{k=k_{inj}} |\Gamma_{\boldsymbol{k}}|^2$ is the constant energy injection forcing, and the caret indicates a spatial Fourier transform. We use a pseudo-spectral method to numerically integrate Eq. S19 in a periodic square domain with each side of length $2\pi$ discretized with $512^2$ collocation points. We choose $\nu_i = 0.5$, $\nu_u = 10^{-35}$, $k_{inj} = 130$ and $f_0 = 0.05$ to generate a statistically stationary turbulent flow with an energy injection rate $\epsilon_{inj} \equiv f_0 / k_{inj}^2 = 2.96 \cdot 10^{-6}$. A representative pseudo-color plot of the vorticity field is shown in Fig. S7**a** and we plot the corresponding energy spectrum in Fig. S7**b**. The energy spectrum shows nearly a decade of $k^{-5/3}$ inverse energy cascade.

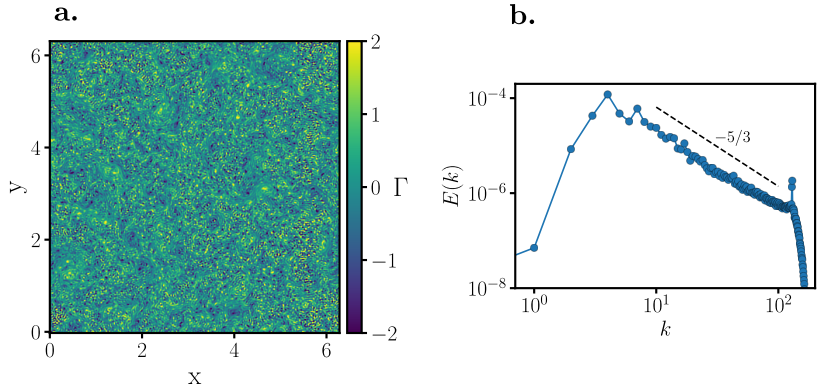

Figure S7: (**a.**) Representative pseudo-color plot of the vorticity field $\Gamma$, and (**b.**) the corresponding energy spectrum. The dashed line indicates the $k^{-5/3}$ scaling.

## 8 Supplementary Movies

The dynamics of swarm for Model A and and Model B is shown in "movie-A-floppy.mp4" and "movie-B-rigid.mp4" respectively. For both movies, the behaviour of the swarm is observed in the presence of background turbulent field with $n = 500$ modes and $L = 20R_s$. For Model A, we use non-affine force strength $h_X = -1000$, turbulent field strength $V_0 = 0.5$, $\gamma = 1$, and $N = 32$ particles. Corresponding values for Model B movie are $h_X = -1, V_0 = 1, \gamma = 1$. In both the movies, for clarity, $\lambda$ is kept constant throughout so that the equations of motion are deterministic.

Similarly, dynamics of the swarm with harmonic interactions is shown in movie "movie-springs.mp4". Here, the stiffness of the harmonic interaction $K = 500$ and the background turbulent field is simulated for the same parameters as in other movies.



\end{document}